\documentclass[twocolumn,prl,ams,superscriptaddress,nofootinbib,showpacs]{revtex4}
\usepackage{graphicx}
\usepackage{epstopdf}
\usepackage{dcolumn}
\usepackage{amssymb,amsmath}
\usepackage{amsfonts}
\usepackage{bbm}
\usepackage{bm}
\usepackage{undertilde}
\newcommand{\AdS}{$\mathrm{AdS}$}
\newcommand{\AdSt}{$\mathrm{AdS}_3$}
\newcommand{\dS}{$\mathrm{dS}$}
\newcommand{\dl}{\text{d}l}

\begin{document}

\title{Conformal Smectics and their Many Metrics}
\author{Gareth P. Alexander}
\affiliation{Centre for Complexity Science and Department of Physics, University of Warwick, Coventry, CV4 7AL, UK}
\author{Randall D. Kamien}  
\affiliation{Department of Physics and Astronomy, University of Pennsylvania, Philadelphia, PA 19104-6396, USA}
\author{Ricardo A. Mosna}
\affiliation{Department of Physics and Astronomy, University of Pennsylvania, Philadelphia, PA 19104-6396, USA}
\affiliation{Instituto de Matem\'atica, Estat\'\i stica e Computa\c{c}\~ao Cient\'\i fica, Universidade Estadual de Campinas, 13083-859, Campinas, SP, Brazil.}
\date{\today}

\begin{abstract}
We establish that equally-spaced smectic configurations enjoy an infinite-dimensional conformal symmetry and show that there is a natural map between them and null hypersurfaces in maximally-symmetric spacetimes. By choosing the appropriate conformal factor it is possible to restore additional symmetries of focal structures only found before for smectics on flat substrates.  
\end{abstract}
\maketitle

Not only do symmetries characterize and constrain the structure of physical theories, they also allow us to choose convenient frames, coordinates, and variables to analyze and formulate our questions.  Symmetries of the ground state manifold are especially interesting, being the deep origin of Nambu-Goldstone modes and the consequential topological defects. Ground states in smectics have broken rotational and translational symmetries which lead to disclinations and dislocations as topological excited states.  However, smectics are easily identified in the laboratory through the formation of focal conic domains -- where defects take the shape of conic sections --, which are a hallmark of layer order~\cite{friedel10}. In prior work \cite{Poincare1} it was found that there exists a hidden symmetry of these focal conic domains, namely they admit a natural action of the Poincar\'e group on a Minkowski spacetime that extends the space on which the smectic lives. In this Letter we show how this formalism extends to describe smectics on curved substrates while retaining the hidden symmetry amongst textures through an infinite dimensional conformal freedom in the choice of spacetime metric. 

Geometry and topology play a prominent role in determining the order and properties of soft materials~\cite{lubensky92,Nelson,bowick09}. Textures that would be suppressed by a large energetic cost in flat space can become energetically preferred, or may even be an unavoidable requirement of topology. For instance, smectics on bumpy surfaces show an accumulation of dislocations in regions of positive Gaussian curvature, and the flat space ground state of straight equally spaced layers is frustrated by focusing of the layer normals and the formation of cusps~\cite{columns,columns2}. On compact spaces, such as the sphere, defects are 
typically a topological necessity and their presence then impacts upon all aspects of the possible textures and their potential modes of relaxation or low energy excitations~\cite{xing}. Though the presence of disclinations and dislocations in smectics on curved substrates has been discussed with ingenious applications of differential geometry~\cite{lubensky92,xing}, and the formation of cusps in response to surface curvature demonstrated to be a generic motif~\cite{columns,columns2}, direct analogs of the exquisite focal conic domains that so typify smectics in flat space have not yet been discussed. 

The insight of \cite{Poincare1} is that seemingly distinct focal conic textures are in fact related in a precise way, through a hidden symmetry revealed by analogy with special relativity. The extension to a curved setting raises the question of whether or not a similar hidden symmetry exists there, perhaps exploiting ideas from general relativity and the structure of curved spacetime, and indeed this turns out to be the case. First let us recall the construction in flat space. 

Smectics are described via level sets of a phase field $\phi({\bf x})=na,\,n\in\mathbb{Z}$, where $a$ is the layer spacing, and are governed by a free energy that penalizes bending, $\kappa\equiv{\bm\nabla}^2\phi$, and layer compression, $e\equiv(1-\vert{\bm\nabla}\phi\vert)$.  The only way to have vanishing curvature and compression is to build equally spaced, planar layers \cite{DiDonnaKamien,Kleman}.  More generally, at long lengthscales, bending energy is much less costly than compression. This favors configurations where the compression vanishes away from a set of defects, which may be idealized as curves and points. Geometric insight into these textures can be gained by considering the surface $S$ given by the graph of $\phi$ inside $\mathbb{R}^{d+1}$ \cite{chen09}. When the compression vanishes these are constant angle surfaces \cite{cermelli07} (in the Euclidean sense) with the surface normal \cite{fn1} $\utilde{N}$ and one basis vector of the tangent plane, $\utilde{T}$, making an angle of $\pi/4$ with the $\phi$-direction. A deeper insight is furnished by a component-by-component bijection of $(d+1)$-tuples between (Euclidean) $\mathbb{R}^{d+1}$ and Minkowski spacetime $\mathbb{M}\equiv\mathbb{R}^{d,1}$. Then with respect to the Minkowski metric $\text{d}s^2=\text{d}\phi^2-\text{d}{\bf x}^2$, both $\utilde{N}$ and $\utilde{T}$ are {\sl null} in $\mathbb{R}^{d,1}$ and the surface $S$ is a null hypersurface. Since $S$ is null, through every regular point passes a unique light ray (null geodesic) of $\mathbb{M}$ that is contained entirely in $S$. These light rays making up $S$ project to a family of geodesics in $\mathbb{R}^d$ whose tangents at every point coincide precisely with the director field ${\bf n}={\bm\nabla}\phi$, the unit normal to the smectic layers. 

\begin{figure}[t]
\centering
\includegraphics[width=.5\textwidth]{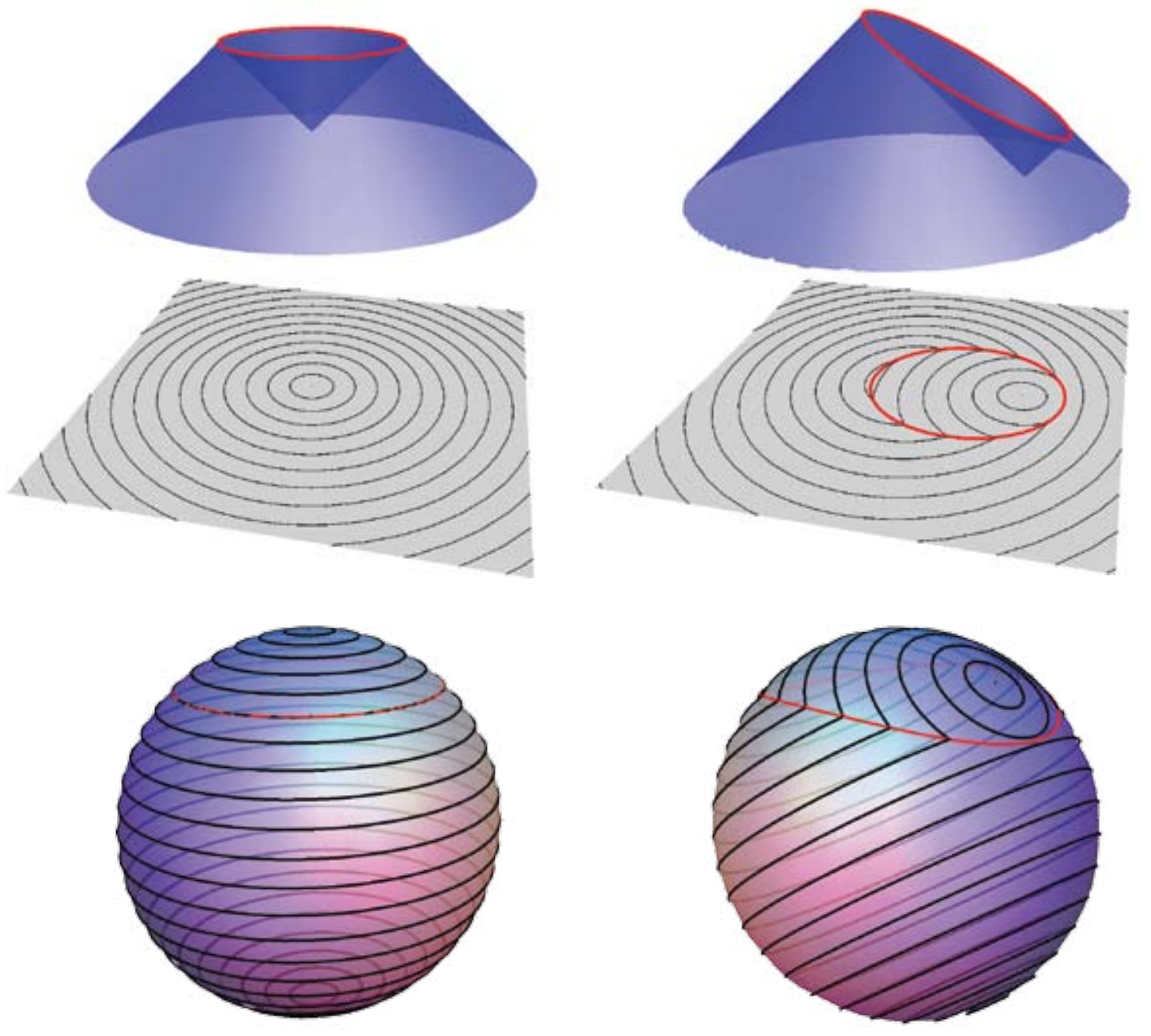}
\caption{Null surfaces, such as the intersection of two light cones, in Minkowski can be used to describe equally spaced smectic textures on the plane, if we adopt the usual Cartesian coordinates for Minkowski (upper panel). By using a different set of coordinates, such as the Carter-Penrose coordinates of \eqref{eq:conformal_Minkowski}, the {\sl same} null surface can also be used to describe equally spaced smectic textures on the sphere (lower panel). Viewing the same null surface from different Lorentz frames (left and right columns) yields different smectic textures, but relates them through the spacetime symmetry.}
\label{fig:ellipse}
\end{figure}

Though this general structure establishes the connection with the more traditional geometric approaches \cite{columns,columns2,Kleman,Kleman2,xing}, the spacetime perspective adds the additional insight that different configurations of layers are connected through the symmetries of Minkowski spacetime. Null surfaces in Minkowski remain null under Lorentz transformations but space and time, {\sl i.e.}, $\phi$, are mixed. Equal time slices of the same null surface using the new time coordinate $\phi^{\prime}$ yields an equally spaced smectic texture that is distinct from the original one, but related to it through the spacetime transformation. For example (FIG.~\ref{fig:ellipse}, upper panel), the surface consisting of the pair of light cones with vertices at $\utilde{\cal P}_1=({\bf x}_1,\phi_1)$ and $\utilde{\cal P}_2=({\bf x}_2,\phi_2)$ has a ``focal'' curve in $\mathbb{M}$ (and a corresponding, experimentally visible, projection onto $\mathbb{R}^2$) where they intersect, which, directly from the definition of conic sections, is a conic. We use the term ``focal curve'' to name any locus of points where the director field is discontinuous and thus visible under standard microscopy.  Under a Lorentz boost, $\phi$ and components of ${\bf x}$ mix but we still have two light cones, just with a new intersection locus. As it turns out there is a symmetry between equally-spaced groundstates on {\sl different} substrates.  

Extending  the general framework to smectics on an arbitrary curved space $\mathbb{U}$ with metric $\dl^2 = h_{ij}\text{d}x^i\text{d}x^j$ is immediate: the correspondence with null surfaces that we have described is quite general and the entire discussion carries over {\sl verbatim}. Equally spaced smectic textures are in direct correspondence with level sets of null hypersurfaces in a spacetime with metric $\text{d}s^2=\text{d}\phi^2-\dl^2$ and the null geodesics which rule the surface $S$ project to geodesics of the so-called optical metric $\dl^2$ on $\mathbb{U}$. What is less immediate is that this correspondence still has any deep insights to offer, and in particular that it continues to reveal any hidden symmetries between smectic textures. The spacetime metric $\text{d}s^2=\text{d}\phi^2-\dl^2$ does not have the symmetries of Minkowski and is not invariant under Poincar\'e transformations. Indeed there are no symmetries that mix the space and time coordinates, for the spacetime has a simple product form, $\mathbb{U}\times\mathbb{R}$, and all the curvature lies in the spatial sections. ``Boosts'' would mix the curved and flat pieces and render a distinct spacetime. With this metric there are no hidden symmetries. Fortunately, however, this is not the only choice we can make. 

The key point is that we are restricting our attention to smectic textures that are everywhere equally spaced and that these correspond to null surfaces in a Lorentzian spacetime. It is well known that the null structure of a spacetime is preserved under conformal rescalings of the metric~\cite{HawkingEllis,fn2}. Thus a given surface, for instance the intersection of two light cones, will appear null in any of the conformal family of metrics
\begin{equation}
\text{d}s^2 = \Omega^2 \Bigl[ \text{d}\phi^2 - h_{ij}({\bf x})\text{d}x^i\text{d}x^j \Bigr] ,
\label{eq:optical_metric}
\end{equation}
where $\Omega(\phi,{\bf x})$ is an arbitrary conformal factor, and {\sl any} member of this conformal family of spacetimes can be used to describe equally spaced smectic textures on $\mathbb{U}$ (with metric $\text{d}l^2$). 
A natural question, then, is what choice of conformal factor should be made? And are there any choices for which the hidden symmetry is restored?  

An example, well known in relativity, provides some immediate insight. The conformal structure of $(2+1)$-dimensional Minkowski spacetime can be revealed by defining new coordinates $(\phi,\alpha,\beta)$ through $t=\Omega_M\sin\phi, x+iy=\Omega_M\sin\alpha\,\text{e}^{i\beta}$ with $\Omega_M=[\cos\phi+\cos\alpha]^{-1}$, producing the metric~\cite{HawkingEllis} 
\begin{equation}
\text{d}s^2_{M} = \Omega_M^2 \Bigl[ \text{d}\phi^2 - \text{d}\alpha^2 - \sin^2\alpha\,\text{d}\beta^2 \Bigr] .
\label{eq:conformal_Minkowski}
\end{equation}
Thus Minkowski is conformal to $\mathbb{S}^2\times\mathbb{R}$. Null surfaces in Minkowski, viewed using the metric \eqref{eq:conformal_Minkowski}, correspond to equally spaced smectic layers on the surface of the two-sphere $\mathbb{S}^2$. The general apparatus that was constructed for relating null surfaces in Minkowski to focal conics in flat space~\cite{Poincare1}, carries over in its entirety to furnish a description of equally spaced smectic textures on the sphere (in any dimension, in fact), simply by switching between the usual Cartesian coordinates $(t,x,y)$ of Minkowski and the Carter-Penrose coordinates $(\phi,\alpha,\beta)$ of the conformal version of the metric \eqref{eq:conformal_Minkowski}. Since the spacetime is Minkowski, we recover the natural action of the Poincar\'e group that mixes space and time coordinates, revealing once again a hidden symmetry between the smectic textures, see FIG.~\ref{fig:ellipse}.

Moreover, the conformal freedom reveals a much greater structure, for not only are seemingly distinct smectic textures on $\mathbb{S}^2$ related via a spacetime symmetry, they are also related to  textures (and symmetries) of smectics in flat space. They are, after all, the same null surface, in the same spacetime, just being viewed by different observers. Similarly, by choosing coordinates $(\phi,\alpha,\beta)$ defined by $t=\text{e}^{\phi}\cosh\alpha, x+iy=\text{e}^{\phi}\sinh\alpha\,\text{e}^{i\beta}$, we cover the interior of the future light cone through the origin in Minkowski with a metric conformal to the standard one on $\mathbb{H}^2\times\mathbb{R}$.
Thus with this choice of coordinates the same null surfaces in Minkowski can be used to describe equally spaced smectic textures on the hyperbolic plane!

Can we go further? Are there other choices of coordinates such that Minkowski is conformal to $\mathbb{U}\times\mathbb{R}$ for an arbitrary spatial surface $\mathbb{U}$? The answer is no: maximally symmetric spacetimes can only accommodate metrics of the form \eqref{eq:optical_metric} with maximally symmetric spatial sections~\cite{walker44}, {\sl i.e.}, $\mathbb{U}$ is either (i) the plane $\mathbb{R}^d$, (ii) the sphere $\mathbb{S}^d$, or (iii) the hyperbolic plane $\mathbb{H}^d$. Other spatial sections require a less symmetric spacetime. 
However, it is also interesting to observe that we need not restrict ourselves to Minkowski as the underlying spacetime: the conformal freedom in \eqref{eq:optical_metric} allows us both to view smectics on different spatial sections with the same spacetime and to view smectics on the same spatial section with different spacetimes. For instance, smectics on each of the three maximally symmetric spaces ($\mathbb{R}^d,\mathbb{S}^d,\mathbb{H}^d$) can be described using an optical metric \eqref{eq:optical_metric} corresponding to any of the three maximally symmetric spacetimes: Minkowski ($\mathbb{M}$), de Sitter (\dS), and anti-de Sitter (\AdS).This then raises the question: why use Minkowski? And, is there a reason to choose one representation over the others?

Certainly for smectic textures in flat space, Minkowski may seem to be the natural choice. But is it still the natural choice for smectics on the sphere? We suggest that here the use of anti-de Sitter has some potential advantages. First, \AdS\ is distinguished among the three by having a periodic time direction and since we wish to associate this to the smectic phase $\phi$ -- itself a periodic quantity -- the use of anti-de Sitter may bring certain benefits, especially if we consider smectic textures with dislocations. A second distinguishing feature of \AdS\ is that it is the only homogeneous spacetime in which we can view space as a sphere and where the time coordinate $\phi$ is associated to a proper Killing vector field, rather than just a conformal one; for the same reason we might choose $\mathbb{M}$ and \dS\ for smectics on $\mathbb{R}^2$ and $\mathbb{H}^2$, respectively. 
Although this is not crucial for the equally spaced textures we consider here, it may prove useful when the compression is non-zero. For these reasons we summarize briefly the use of \AdS\ in describing smectic textures on $\mathbb{S}^2$. Although a complete classification of all possible textures is challenging, for the same reasons as described in \cite{Poincare1} a very large class of textures, covering all experimentally observed focal conic textures, may be pieced together from the intersections of null planes and light cones. Thus it suffices to describe the generic features of these simple null surfaces, from which any desired texture can be constructed. Note that this construction works in any dimension, but we will focus on the relevant case of two-dimensional smectics.

Recall that \AdS$_{d+1}$ can be isometrically embedded in $\mathbb{R}^{d,2}$, with its natural metric, as the hyperboloid 
\begin{equation}\label{eq:AdS}
t_1^2 + t_2^2 -{\bf x}^2 = 1 .
\end{equation}
The isometries of \AdS\ are then generated by the action of $SO(d,2)$ on this hyperboloid and they descend to a natural action of $SO(d,2)$ on the set of smectic textures on $\mathbb{S}^d$. The choice of coordinates $t_1+it_2=\sec\alpha\;\text{e}^{i\phi}$, $x_1+ix_2=\tan\alpha\;\text{e}^{i\beta}$ in $\mathbb{R}^{2,2}$ provides a parameterization of \AdSt\ such that the induced metric is of the form \eqref{eq:optical_metric} with $\Omega=\sec\alpha$ and where the spatial metric $\text{d}l^2$ is the usual round metric on $\mathbb{S}^2$. The coordinates $\phi,\alpha,\beta$ cover the entire \AdSt\ with $\phi\in[0,2\pi),\,\alpha\in[0,\pi/2),\,\beta\in[0,2\pi)$ and thus only cover one hemisphere of the $\mathbb{S}^2$, with the conformal factor $\sec\alpha$ taking the equator to infinity in \AdSt.
\begin{figure}[t]
\centering
{\includegraphics[width=.4\textwidth]{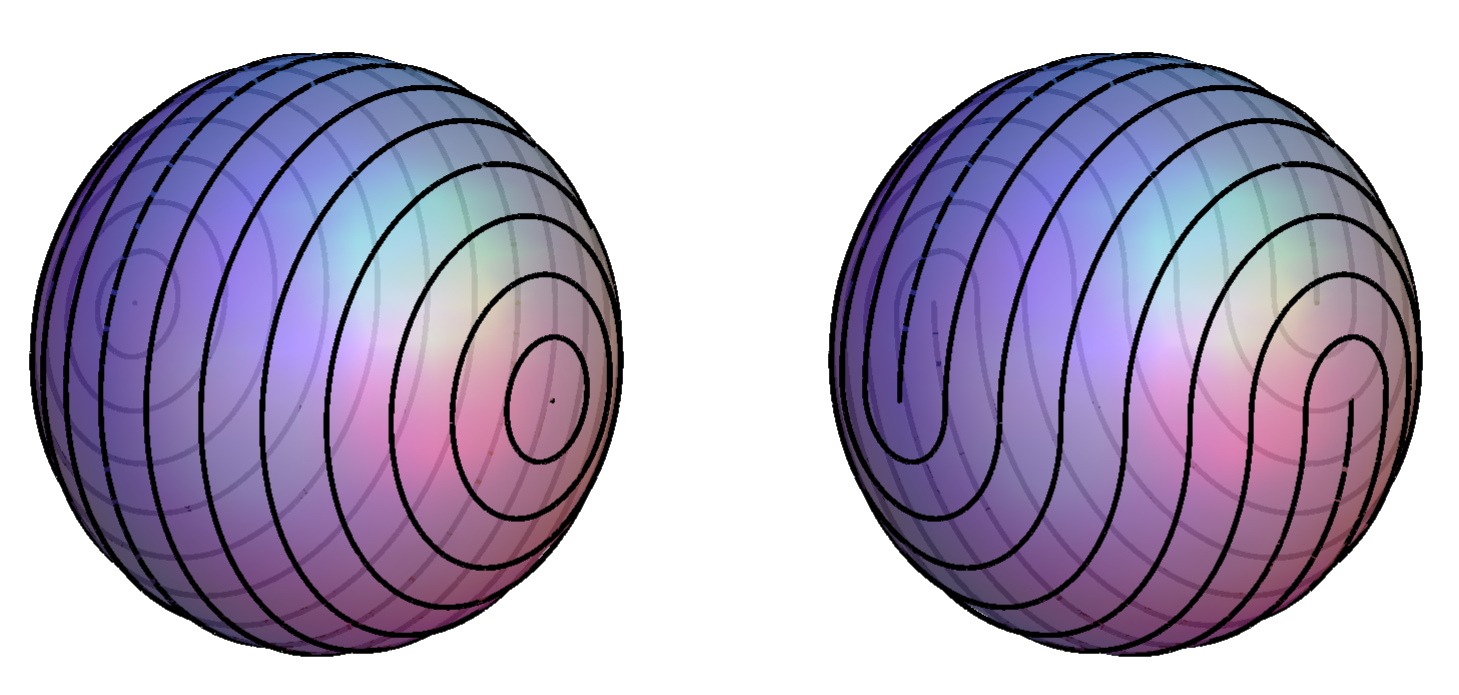}}
\caption{Smectic texture on a sphere derived from a pair of null `planes' in \AdSt. Left: the two planes have the same orientation. Right: the two planes have orthogonal orientations. See text for details. }
\label{fig:sphere_planes}
\end{figure}

From the point of view of smectics, the advantage of embedding in $\mathbb{R}^{2,2}$ is that null surfaces in \AdSt\ are also null in $\mathbb{R}^{2,2}$, which provides a convenient means of constructing them. Starting with the analog of smectic ground states, we can take as a representative null plane in \AdSt\ the intersection of the hyperboloid \eqref{eq:AdS} with the plane $x_1=t_1$. In terms of our coordinates $(\alpha,\beta,\phi)$ this is the relation $\sin\alpha\cos\beta=\cos\phi$ and thus equally spaced values of $\phi$ correspond to equally spaced layers $x=\text{const.}$ on the standard embedding of the sphere into $\mathbb{R}^3$, $x^2+y^2+z^2=1$. Now, the plane $x_1=t_1$ intersects \AdSt\ in two disjoint pieces, $t_2\geq 1$ and $t_2\leq -1$, so that we are really describing two null planes in \AdSt. However, as the coordinates $\alpha,\beta$ only cover one hemisphere, this duplicity is useful in defining the smectic texture on the entire $\mathbb{S}^2$: the two pieces can be mapped to different hemispheres and glued together along the equator, corresponding to infinity in \AdSt. Although this process leads to $+1$ point defects on the equator of the sphere, these defects are nowhere to be found in \AdSt\ since it does not include the equator. More generally, when gluing together the two hemispheres, there is no reason for us to choose the same orientation of plane waves: for instance, we may take the second plane wave to be given by the intersection of the plane $t_1=x_2$ with \AdSt\ ($t_2\leq -1$), as shown in FIG. \ref{fig:sphere_planes}. When they are different, there are four $+1/2$ point defects on the equator, but otherwise the layers can be made to join so that the normal is continuous. 

Now we consider lightcones in \AdSt\ and their related disclinations in $\mathbb{S}^2$. Since \AdSt\ is maximally symmetric, its geometry looks exactly the same everywhere. A lightcone at $\utilde{\cal P}=(1,0,0,0)$ gives rise to a point defect at the North pole of $\mathbb{S}^2$, with lines of constant latitude as layers. Rotations in $t_1\wedge t_2$ and $x_1\wedge x_2$ shift $\phi$ and $\beta$, respectively. The former evolve time and the latter just reparametrize the layers. Boosts in $t_2\wedge x_1$ and $t_2\wedge x_2$ fix $\utilde{\cal P}$ and thus also just reparameterize the layers.  However, a boost in $t_1\wedge x_1$ with velocity $v=\sin\psi$ maps the condition $t_1=1$ to $t_1 \sec\psi-x_1\tan\psi=1$, which is equivalent to $\cos\phi={\bf s}\cdot{\bf n}$, where ${\bf n}=(\sin\psi,0,\cos\psi)$ and ${\bf s}=(\sin\alpha\cos\beta,\sin\alpha\sin\beta,\cos\alpha)$ is an arbitrary point on the sphere. As a result, layers of constant latitude are rotated around the axis $\hat{\bf y}$ by an angle $\psi$. Similarly, a boost in $t_1\wedge x_2$ yields a rotation around $\hat{\bf x}$.

We finally come to the question of focal sets. Consider a point $({\bf s},\phi)$ at the intersection between the future lightcone emanating from $({\bf f}_1,\phi_1)$ and the past lightcone emanating from $({\bf f}_2,\phi_2)$. Denoting the distance along the sphere by $\mathrm{d}$, this means that $\mathrm{d}({\bf s},{\bf f}_1)=\phi-\phi_1$ and $\mathrm{d}({\bf s},{\bf f}_2)=\phi_2-\phi$ so that  $\mathrm{d}({\bf s},{\bf f}_1)+\mathrm{d}({\bf s},{\bf f}_2)=\phi_2-\phi_1$, the equation for an ellipse. Similarly, the intersection of two future or two past lightcones yields $|\mathrm{d}({\bf s},{\bf f}_1)-\mathrm{d}({\bf s},{\bf f}_2)|=|\phi_1-\phi_2|$, the equation for a hyperbola. Note, however, that the distinction between ellipses and hyperbol\ae\ is artificial on the sphere, since a lightcone always refocuses after a time $\pi$; the equation for a hyperbola with foci at ${\bf f}_1,{\bf f}_2$ goes into the equation for an ellipse with a foci at ${\tilde{\bf f}}_1,{\bf f}_2$, with ${\tilde{\bf f}}_1$ the antipodal point of ${\bf f}_1$. In fact, even a parabola on the sphere is an ellipse: the locus of points equidistant from an arbitrary point and a great circle is identical to the locus generated by either an ellipse or hyperbola between the arbitrary point and the conjugate poles of the great circle.  

In closing, we note that whenever the focal curve is an ellipse on any space, then simple geometry allows us to see that the energy of a simple focal domain arises only from the focal curve!  Consider the focal set depicted on the right in FIG. 1.  The null surface only differs from that on the left along the cusp: cutting along the rim and flipping the well over results in the null surface for a single disclination.  Since no bending or stretching were necessary it follows that the the only energy difference between the right and the left is concentrated on the focal curve.  Thus, though the elastic energy for a smectic
\begin{equation}
F=\frac{1}{2}\int \text{d}^d\!x\left\{ B\left[\left(\bm{\nabla}\phi\right)^2-1\right]^2 + K\left(\bm{\nabla}^2\phi\right)^2\right\}
\end{equation}
is not invariant under the hidden Poincar\'e, $SO(2,2)$, or $SO(3,1)$ symmetries of the underlying space, the energy of these focal domains transforms {\sl locally}.  The compression energy remains vanishing under such transformations and the bending energy is only different on the focal curve.  This geometric transformation extends to all ellipsoidal focal sets in three dimensions, and so forth.  

It does not appear, however, that any such simplification occurs for more complex domains, in particular the toric focal conic domains.  Though  the {\sl special conformal transformation}  of Minkowski space can be used to map a circular focal set to any other conic \cite{note5}, there exists an immediate obstruction to any sort of energetic comparison because the hyperbolic and parabolic focal sets are not compact.  Either they run off to infinity or they end on point disclinations \cite{Poincare1}.  Once we admit focal curves we must also admit point defects; once we admit point defects we are compelled to set boundary conditions at infinity to conserve topology or, as is the usual case, we can add the point at infinity and study our problem on the sphere. Thus our ability to study smectics on compact surfaces such as $\mathbb{S}^2$ becomes necessary to properly formulate these problems.

We note that the conformal freedom of the null hypersurfaces is a sort of pointwise realization of the projective geometry of light cones.  Whether we can extend these ideas or exploit the full power of Lie sphere geometry \cite{Lie} to understand the geometric symmetry of the full energy or the structure of higher genus spaces remains open.

In future work we will study these issues, generalizations to higher dimensions, and to more complex focal structures.  In addition we will explore the use of \AdS\ to study the inclusion of dislocations and smectic textures with non-uniform spacing.
 
% acknowledgements
It is a pleasure to acknowledge discussions with D. Beller, B.G. Chen, and M. Trodden and, in particular, a mind-blowing discussion with G. Jungman. RAM acknowledges financial support from FAPESP. This work was supported in part by NSF Grant DMR05-47230.


\begin{thebibliography}{1}
\bibitem{fn1} Bold symbols are always vectors in $\mathbb{R}^d$; (d+1)-vectors will be denoted as $\utilde{V}=({\bf v},v^{d+1})$.  Inner products and lengths are calculated with respect to the ambient metric; when it is unclear by context, we will be explicit.  Greek indices are used in $\mathbb{R}^{d,1}$, Roman in $\mathbb{R}^d$.
\bibitem{fn2} Let $f(\utilde{x},\utilde{\dot x}) = \frac{ds}{d\lambda}$.  Extremizing $\int \Omega\text{d}s=\int\Omega f \text{d}\lambda$ gives the geodesic equation:
$0=\Omega(-\frac{d}{dt}\partial_{\utilde{\dot x}}f + \partial_{\utilde{x}}f) + \partial_{\utilde{x}}\Omega f$.  Since $f=0$ for null geodesics, null geodesics are the same in both metrics.
\bibitem{note5} It is amusing to note that in Minkowski space {\sl all} the focal conic sections can be obtained via the {\sl special conformal transformation},
$x^\mu \rightarrow (x^\mu - b^\mu \utilde{x}^2)/({1-2b_\mu x^\mu + \utilde{b}^2 \utilde{x}^2})$,
of the simplest circular focal set: consider the two timelike separated lightcones in $\mathbb{R}^{2,1}$ with vertices at $\utilde{\cal P}_1=(0,0,0)$ and
$\utilde{\cal P}_2=(0,0,2t)$.  These two cones intersect on a circle of radius $t$ at time $t$: $\Sigma=\{(t\cos\theta,t\sin\theta,t)\vert \theta\in[0,2\pi) \}$, which casts a focal circle on the $xy$-plane.  Taking $\utilde{b}=(b,0,0)$, $\Sigma$ becomes $\Sigma'=\{(t\cos\theta,t\sin\theta,t)/(1-2b\cos\theta)\vert \theta\in[0,2\pi) \}$ which gives a planar curve in polar coordinates 
$
r(\theta) = t/(1-2bt\cos\theta),
$
the general equation for a conic with eccentricity $2bt$ and {\sl latus rectum} $2t$.  Were $\utilde{b}=(0,0,\beta)$ timelike, the original circle of radius $t$ would become a new circle of radius $t/(1-\beta t)$.
\end{thebibliography}
\end{document}